\documentclass[onecolumn,draftclsnofoot,12pt]{IEEEtran}
\usepackage[T1]{fontenc}
\usepackage[latin9]{inputenc}
\usepackage{graphicx}
\usepackage[unicode=true,
 bookmarks=true,bookmarksnumbered=true,bookmarksopen=true,bookmarksopenlevel=1,
 breaklinks=false,pdfborder={0 0 0},pdfborderstyle={},backref=false,colorlinks=false]
 {hyperref}
\hypersetup{pdftitle={Your Title},
 pdfauthor={Your Name},
 pdfpagelayout=OneColumn, pdfnewwindow=true, pdfstartview=XYZ, plainpages=false}
\usepackage{breakurl}

\makeatletter
 \let\oldforeign@language\foreign@language
 \DeclareRobustCommand{\foreign@language}[1]{%
   \lowercase{\oldforeign@language{#1}}}

\usepackage[caption=false,font=footnotesize]{subfig}

\makeatother

\begin{document}

\title{Cell-less Communications in 5G Vehicular Networks Based on Vehicle-Installed
Access Points}

\author{Lijun~Wang, Tao~Han, Qiang~Li, Jia~Yan, Xiong~Liu, and~Dexiang~Deng\thanks{Accepted by IEEE Wireless Communications SI on ``Emerging Technology
for 5G Enabled Vehicular Networks''.}\thanks{Lijun~Wang is with Wuhan University and Wenhua College;}\thanks{Tao~Han, Qiang~Li, and Xiong~Liu are with Huazhong University of
Science and Technology;}\thanks{Jia~Yan and Dexiang~Deng are with Wuhan University.}\thanks{The corresponding author is Tao Han.}\thanks{Copyright (c) 2017 IEEE. Personal use of this material is permitted.
However, permission to use this material for any other purposes must
be obtained from the IEEE by sending a request to pubs-permissions@ieee.org.}\thanks{Digital Object Identifier: 10.1109/MWC.2017.1600401}}

\markboth{IEEE Wireless Communications, vol. 24, no. 6, pp. 64\textendash 71,
Dec. 2017.}{}
\maketitle
\begin{abstract}
The development of intelligent transportation systems raises many
requirements to the current vehicular networks. For instance, to ensure
secure communications between vehicles, low latency, high connectivity
and high data rate are required for vehicular networks. To meet such
requirements, the 5G communication systems for vehicular networks
should be improved accordingly. This article proposes a communication
scheme for 5G vehicular networks, in which moving access points are
deployed on vehicles to facilitate the access of vehicle users. Moreover,
the adjacent vehicle-installed moving access points may cooperatively
communicate with the vehicle users by joint transmissions and receptions.
In this way, the vehicle users communicate with one or more unspecified
cooperative access points in a cell-less manner instead of being associated
to a single access point. To manage such cell-less networks, local
moving software-defined cloudlets are adopted to perform transmission
and scheduling management. Simulation results show that the proposed
scheme significantly reduces the latency, while improving the connectivity
of the vehicular networks, and can be considered as a research direction
for the solution to 5G vehicular networks.
\end{abstract}

\newpage{}

\section{Introduction}

For the safety of driving, the drivers or even the auto-pilot controllers
have to monitor the real-time vehicle and traffic status. The most
important way to obtain such vehicle and traffic status is receiving
the information from other vehicles and public services by the vehicular
networks (VNETs). This driving-related information includes the traffic
status information, safe driving and vehicle fault diagnosis information,
and navigation information from the service provider. Besides the
mentioned critical real-time information for driving, the entertainment
information, information of common communication services, and social
information also need to be transmitted by the VNETs \cite{amadeo_information-centric_2016_IEEECommun.Mag.}.
Considering the application prospects of VNETs, it is necessary to
construct a universal and powerful communication platform for VNETs.
It is feasible to deploy application services and network equipment
close to the moving vehicles, such that the latency of network transmission
and service response time can be reduced significantly. One of the
important approaches to providing services close to the vehicles is
the mobile edge computing scheme. By mobile edge computing, computing
and storage resources are deployed at a variety of the edge network
equipment to provide rapid services for mobile users \cite{liu_mobile_2017}.
Because of the high mobility of vehicles, the traditional mobile computing
encounters challenges of efficient and rapid resource scheduling and
allocation. Meanwhile, the approach to providing communication services
close to the vehicles is to implement the access network among the
vehicles themselves. Based on this as-close-as-possible deployment
strategy, dynamic, open, self-organized, easy-deploying, and cost-effective
VNETs can be realized instead of traditional monitoring-focused traffic
assisting systems. Consequently, the study on new architectures of
VNET is vital for the future development of intelligent transportation
systems (ITSs).

The concept of VNET grows out of an application of wireless sensor
networks (WSNs), and in the early days of VNETs, many devices and
protocols directly came from WSNs and mobile ad hoc networks (MANETs).
However, because of the quick movement of vehicles, many existing
protocols including ZigBee and Bluetooth are not suitable for VNETs
anymore. In contrast, IEEE 802.11p and Long-Term Evolution (LTE) are
recognized as two most important technologies for VNETs \cite{zheng_heterogeneous_2015_IEEECommun.Surv.Tutor.}.
Based on these technologies, various architectures for VNETs have
been developed. Generally, the different architectures can be classified
into two kinds, that is, the infrastructure-based architectures and
the MANET-based architectures. In the infrastructure-based architectures,
base stations (BSs) or access points (APs) are deployed along the
road, acting as road-side units (RSUs) to provide access services
for the vehicle users \cite{salahuddin_software-defined_2015}. On
the other side, in MANET-based architectures, vehicle users communicate
with each other in a point-to-point manner. Because of the ad-hoc
nature of the MANETs, it is crucial to perform routing and resource
allocation, for which many technologies and solutions have been proposed.
One of the promising approaches is to implement a locally centralized
routing and resource allocation scheme based on software defined mobile
cloudlets \cite{liao_software_2015_MobileNetwAppl}. In the future
5G networks, it is possible to use both infrastructure-based and point-to-point-based
communication modes. We can combine the traditional cellular access
technology and device-to-device (D2D) technology to construct the
future VNETs. It becomes imperative for the studies using these technologies
to provide cost-effective, easily-implementable schemes for the new
generation of VNETs. The schemes should satisfy the demands for low
latency, high connectivity and a large amount of data on the basis
of vehicular secure communications. Moreover, the crucial issue on
how to efficiently connect VNETs to current and future mobile communication
systems remains an open problem.

Both IEEE 802.11p and cellular networks are widely used in VNETs,
and many research works are focusing on them, \cite{vinel_3gpp_2012_IEEEWirel.Commun.Lett.}
gives a theoretical framework to analyze the performance of VNETs
based on 802.11p and LTE, respectively, and concludes that the abilities
of current cellular networks to support secure vehicular communications
are insufficient. On the other hand, considering that the future 5G
mobile communications can provide very attractive quality-of-service
(QoS) regarding delay, data rate, and reliability compared to current
LTE, 5G mobile communication technology becomes a very competitive
candidate for the implementation of VNETs \cite{yu_optimal_2016_IEEETrans.Veh.Technol.}.
Moreover, 5G communication-based VNETs can connect to universal mobile
communication systems at very low costs. 

However, merely using the fixed BSs of 5G communication networks in
VNETs may encounter many issues, including frequent handovers, unsatisfactory
connectivity, and unpredictable latency, which are caused by the mobility
of the vehicle users. Many approaches have been proposed to solve
these issues, and some of them are moving relay or moving AP related.
Among this kind of approaches, there are two typical application scenarios.
The first scenario is about public transportation system, such as
high-speed train (HST), buses, etc. With one or more moving relays
installed on an HST or bus, the mobile terminals on the HST or bus
can communicate with the relays, so as to get access to the core network
\cite{laiyemo_transmission_2016}. In some of these works, moving
APs are used instead of moving relays. In this way, the moving small
cells on the HSTs or buses provide higher access QoS than fixed cells
along the road, because the frequent handovers are avoided. The performances
of these moving AP based schemes greatly depend on the performance
of backhaul links between the moving APs and the core network, which
is investigated in \cite{jangsher_backhaul_2016}. The second scenario
related to moving relays or moving APs is about relays or APs installed
on vehicles. In this scenario, users on other vehicles or even pedestrians
connect to the core network through the moving relays or APs \cite{patra_improving_2016}.
Because of the relative movements between the relays or APs and the
users, the communications also suffer from unavoidable handovers.
Moreover, like the first scenario, the second scenario also faces
the issues of the performance limited by the backhaul links between
moving relays or APs and the core network. In fact, even in fixed
BS based 5G communication networks with mobile users, handover-related
issues are critical factors of the system performance. 

To overcome handover-related performance issues in 5G networks, some
researchers present solutions using cooperative communications between
BSs or APs. Our previous work presents a cell-less communication architecture
based on cooperative communications \cite{han_5g_2016_IEEECommun.Mag.}.
Different from the conventional cellular architecture, a user does
not associate to a single BS or AP. Instead, it may communicate with
cooperative BSs or APs by coordinated multipoint (CoMP) transmissions
and receptions. The software defined networking controller is deployed
to schedule the traffic and allocate the resource globally. In this
way, the cell-less scheme improves the connectivity and reduces the
latency caused by handovers. For both cooperative moving relay and
moving AP schemes, how to implement the cooperative communications
among moving relays or APs, and how to select some of them to cooperate
are crucial issues in the cooperative-communication-based VNET. \cite{feteiha_enabling_2015}
proposes a cooperative communication relaying scheme. In this scheme,
abundant vehicles in urban area cooperatively act as relays to provide
pedestrians access services. The authors use signal-to-noise ratio
(SNR) based relay selection method in the cooperative communications
and analyze the performance. However, considering the critical requirements
of communication latency and connectivity in VNETs, it is still a
significant issue on how the cooperative vehicle-installed APs provide
access for other vehicle users on the road.

In this article, accounting for the frequent handover and outage issues
caused by the high-speed movement of vehicles, we discuss the 5G-communication-based
VNET access technology and applications. We propose a moving-AP-based
5G cell-less VNET scheme, in which, three simple yet workable strategies
are given for selecting vehicles as the cooperative moving APs to
construct a 5G cell-less moving access network. In simulation results,
we present and compare the performance on the connectivity and latency
of various moving AP selection strategies for an illustrative scenario.

\section{From Fixed Cellular Communications to Moving-AP-Based Cell-less Communications}

Currently, one of the important roles of VNETs is to transmit various
kinds of traffic security information, such as the vehicular status,
nearby vehicular type, vehicular moving and traffic disaster. When
transmitting the messages carrying such information, the latency is
a very critical factor, and delay or loss of the messages will cause
severe damages potentially as well. Therefore stringent latency and
connectivity requirements are indispensable in 5G VNETs. In such scenarios,
using traditional cellular networks with D2D communication to support
VNETs may lead to some issues including frequent handovers, high latency,
poor connectivity, and unbalanced load. Among these issues, latency
issue is a particularly critical one for VNETs, and the handover issue
and connectivity issue can also lead to increase the latency. Moreover,
at the intermediate relays in the communication path of a D2D multi-hop
relay communication system, it takes non-negligible time to receive,
process, and transmit data. Thus the more hops there are in the communication
path, the larger latency in VNETs will be resulted in \cite{fodor_overview_2016_IEEEAccess}.

\begin{figure}[tbh]
\begin{centering}
\includegraphics[width=0.95\textwidth]{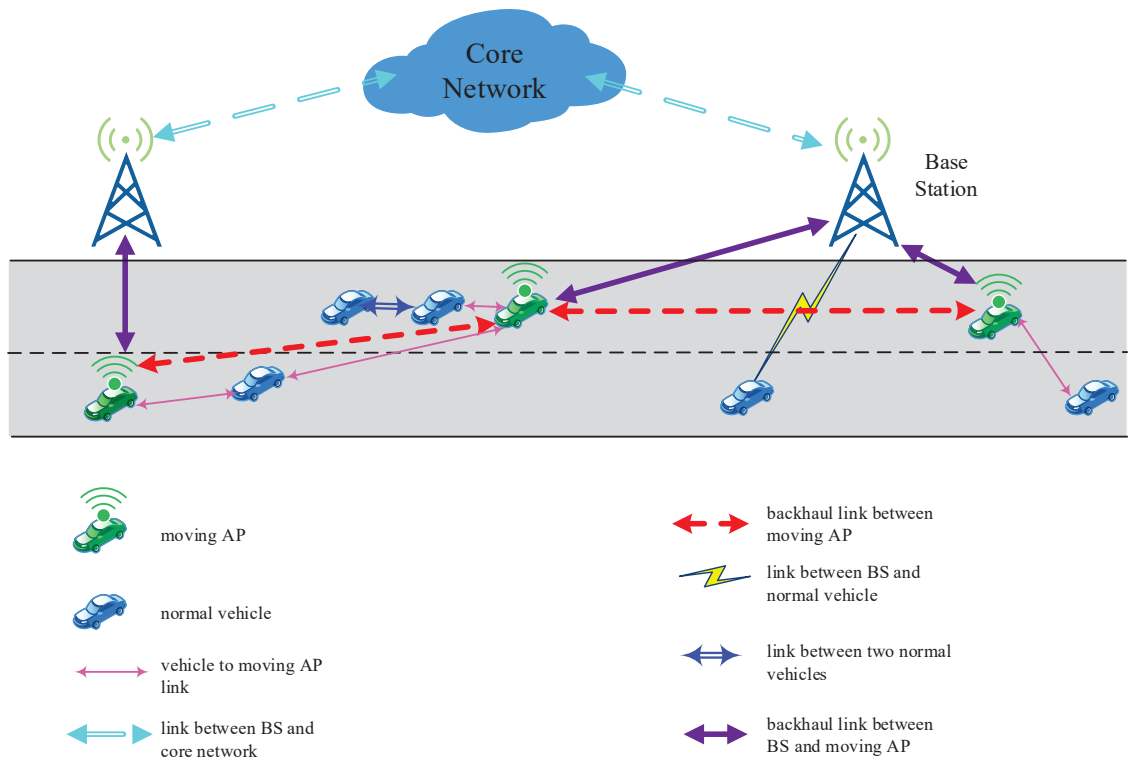}
\par\end{centering}
\caption{An illustration of 5G moving-AP-based cell-less VNET\label{fig:Scenario}}
\end{figure}

There will be a significant challenge to overcome latency issue caused
by the mobility of the vehicles if 5G technologies are directly applied
in VNETs. To solve this problem, as shown in Figure \ref{fig:Scenario},
we propose a scheme to deploy vehicle-installed moving APs on some
selected vehicles. Some of them can be vehicles on which APs are deployed
beforehand, while others may be vehicles with the vehicle-installed
transceivers which can increase their transmission powers and reception
sensitivity by some methods. For example, millimeter wave transmission
may be a key technology in the future 5G networks, which reduces a
single antenna to a millimeter size due to its very short wavelength,
and makes it possible to deploy a massive multi-input multi-output
(MIMO) transceiver on a vehicle. Considering parts of the antennas
of massive-MIMO work when the vehicle acts a regular vehicle user,
we can activate more backup antennas of the massive-MIMO system if
necessary, so that the vehicle can serve as a moving AP as well. The
transmission powers of the vehicles that are not selected as moving
APs remain unchanged to avoid increasing the interference. Considering
the insufficient signal strength at the locations far away from the
moving APs or serious interference in the edge area between two adjacent
moving APs, we propose that, adjacent APs can cooperatively communicate
with vehicle users by joint transmission and reception to provide
vehicle users access services. In this way, there is no \emph{cell}
of the coverage area for any single moving AP, where they cooperatively
communicate to vehicular users in a similar \emph{cell-less} way as
described in \cite{han_5g_2016_IEEECommun.Mag.}. Some advantages
of this access scheme are listed below.
\begin{enumerate}
\item Fewer handovers and outages. When the selected moving APs move together
with the vehicle users at the similar speed along the road, it is
possible to maintain the available links between them for longer durations
than between vehicle users and fixed BSs, thus reducing frequent handovers
and outages. To be specific, most of the handovers will be soft handovers
due to cooperative communications, and other unavoidable handovers
will be replaced by D2D communications between vehicle users or accesses
to fixed 5G BSs.
\item Lower latency. Compared to accessing to fixed 5G BSs or D2D multi-hop
communications that require extra time, there will be much less latency
when vehicle users in a local area communicate with each other by
accessing to nearby moving APs.
\item Better connectivity. 5G moving-AP-based cell-less accessing can provide
better connectivity than fixed-BS-based accessing or D2D-based multihop
transmissions for two reasons. That is, accessing to fixed BSs will
cause frequent handovers of vehicle users, whereas multihop D2D communications
may increase the outage probability while increasing the hops of transmissions.
\item More balanced loads. Because the moving APs are selected from vehicles
heading in the same direction on the road, the density of moving APs
will increase along with an increase in the density of vehicular users
in the congested road. Then the data traffic loads at moving APs will
be more balanced than accessing to fixed BSs.
\end{enumerate}
In the scheme above, to perform joint transmission and reception,
one of the critical factors is how to build the backhaul links between
adjacent moving APs \cite{ge_5g_2014_IEEENetwork}. A feasible approach
is to allocate more bandwidth to APs for backhaul links than conventional
cellular network links and D2D links. Moreover, besides the traditional
cellular microwave transmissions, the emerging 5G transmission technologies
including millimeter wave transmission, massive MIMO and visible light
communication (VLC) can also be utilized to support the backhaul links
for moving APs. Generally, VLC technology can be used for the backhaul
links between moving APs, when they are in the line of sight of each
others moving on the road. However, for the backhaul links between
the core network and moving APs, it is better to use traditional radio
technologies.

\section{Architecture and Modeling of 5G Moving-AP-based Cell-less Communications
in VNETs}

\subsection{Architecture of 5G Moving-AP-based Cell-less Communications}

The architecture of the proposed 5G moving-AP-based cell-less communications
consists of three tiers, that is, the vehicle user tier, the moving
AP tier, and the core network tier. Communications can occur between
tiers or within a tier.

\begin{figure}[tbh]
\begin{centering}
\includegraphics[width=0.8\textwidth]{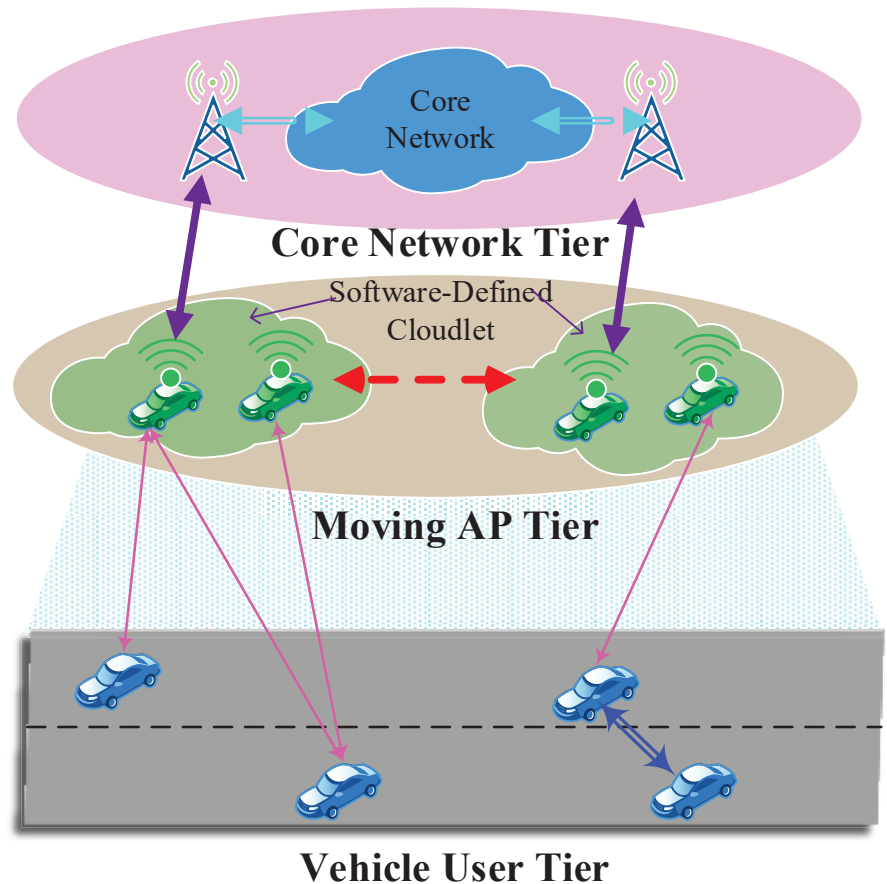}
\par\end{centering}
\caption{Architecture of 5G cell-less VNETs based on moving APs\label{fig:Architecture}}
\end{figure}

As shown in Figure \ref{fig:Architecture}, in the vehicle user tier,
vehicular users can communicate with each other by D2D communications,
especially when they are unable to access to a nearby vehicle-installed
AP or cooperative APs. Moreover, if they cannot communicate with adjacent
vehicle users by D2D links, or there are too many relay hops in the
transmission path of D2D links, then they may try to access to the
nearby or road-side 5G BSs that generally have larger coverage areas.

Generally, a common vehicle user accesses to the nearest moving AP
if the communication link meets its communication demand. When accessing
to a single moving AP cannot provide enough data rate or QoS, a vehicle
user will communicate with its adjacent moving APs which can cooperatively
communicate with it by joint transmission and reception. This cooperative
communication can decrease the co-channel interference from moving
APs and improve signal-to-interference-plus-noise ratio (SINR), thereby
improving the data rate and expanding the coverage area. If the channel
status is even worse and the vehicle users cannot access to cooperative
moving APs, it will try to use D2D links to establish a multi-hop
path to its communication peer. Eventually, there always remains the
choice to access to the nearby 5G BSs, in case that the multi-hop
D2D link may also lead to a considerable latency.

Within the medium moving AP tier, the backhaul links need to be established
among the cooperating APs. In case that it is hard to establish stable
backhaul links with sufficient data rate among the moving APs, backhaul
links between the 5G fixed BSs and moving APs will be used to transmit
control information and data for the cooperative communications. For
the coordination and scheduling management of backhaul links among
the moving APs, local software-defined cloudlets need to be deployed
to manage the moving APs in a particular local area. To be specific,
the candidate cloudlet controllers are deployed together with some
of the moving APs, and among them, the effective cloudlet controllers
will be selected dynamically.

The 5G core network tier consists of the backbone network and the
traditional fixed BSs that are connected to the backbone network.
To support the proposed moving-AP-based scheme, the mobility management
entity of the 5G network should be responsible for the mobility management
of both the mobile user terminals and the moving APs. The cloudlet
controllers in the medium moving AP tier will play a vital role to
help the mobility management entity of the core network to oversee
the mobility in the whole network.

There are several approaches for those vehicles acting as moving APs
to transmit their own information. One of the approaches is to transmit
their information by using extra transceivers if they have. Another
one is a virtual software-only approach, in which a virtual vehicle
user terminal is implemented inside the AP to always access the AP
itself for information transmission.

\subsection{Transmission Modeling for Various Cases of Peer-to-peer Communications}

\begin{figure}[tbh]
\begin{centering}
\includegraphics[width=0.95\textwidth]{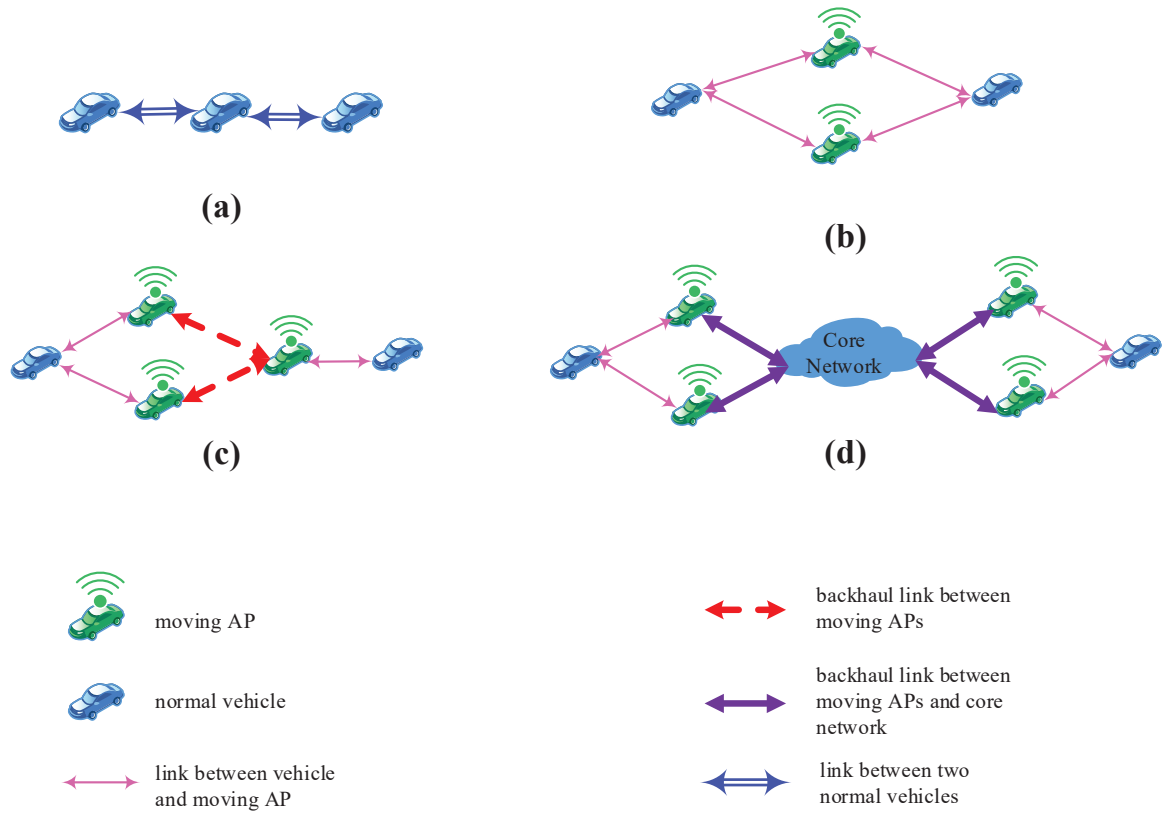}
\par\end{centering}
\caption{Illustrations of transmission paths\label{fig:transmission_path}}
\end{figure}

Vehicle users usually transmit information to other vehicle users
not far away from them, for example, vehicle security information
shall be forwarded among vehicle users within a local area typically.
It means that the communication peers are only a few of hops away
from each other, even if there is no direct link between them. In
5G cell-less VNET based on vehicle-installed moving APs, the transmissions
in peer-to-peer communications can be typically modeled as the several
scenarios, as shown in Figure \ref{fig:transmission_path}.

The case (a) in Figure \ref{fig:transmission_path} shows the D2D-based
communication path, in which there can be one or more hops between
the communication peers. The case (b) shows the scenario of the moving-AP-based
transmissions, in which the communication peers access to a single
AP or a group of cooperative APs, and it requires two hops for the
communication peers to reach each other. When the communication peers
are sufficiently far away from each other, they have to access to
separate APs, between which backhaul links are built to transmit information
and it requires three hops for the communication peers to reach each
other, as shown in case (c). Furthermore, when it is impossible to
establish backhaul links between two accessing APs or two groups of
cooperative APs, the APs transmit information via the 5G core networks
by the backhaul links between the cooperative moving APs and the core
networks. It will result in more hops in the transmission path, as
shown in case (d) in Figure \ref{fig:transmission_path}.

\section{Connectivity and Latency}

\subsection{Selection Scheme of Moving APs}

For analyzing the performance of the proposed 5G cell-less VNET scheme,
it is necessary to determine the spatial distribution characteristic
of the vehicle-installed moving APs, which is indeed decided by the
moving AP selection methods employed. The moving AP selection schemes
can be classified into three kinds, including the predefined selection,
independent random selection, and cooperative selection schemes.

The simplest moving AP selection scheme is the predefined moving AP
selection scheme. By this selection scheme, moving APs are deployed
on some of the intentionally selected vehicles beforehand. Compared
to the ordinary vehicle users, the moving APs should have higher transmission
power and reception sensitivity, which can be achieved by deploying
powerful on-board units on some of the vehicles. For simplicity in
performance analysis, the pre-deployed moving APs can be regarded
as independent and randomly distributed in the vehicles. Then it is
reasonable to assume that the moving APs also follow a Poisson point
process. The predefined moving AP selection method is easy to implement,
and there is no overhead for selection in run-time. However, this
predefined selection method cannot adapt to the constantly changing
traffic situation and communication demands.

Compared to the predefined selection scheme, the independent random
selection scheme and the cooperative selection scheme can adapt to
the actual traffic situation. In the independent random selection
scheme, vehicles need not exchange information with one another for
the selection. According to a certain probability, every vehicle that
is capable of acting as a moving AP decides independently to be a
moving AP or not. This selection scheme is straightforward and easy
to implement, and it does not require additional communication and
computing overhead. Similar to the predefined selection scheme, for
simplicity, we can assume the candidate vehicles follow a Poisson
point process in the performance analysis. In this way, it leads to
a thinned Poisson point process for the distribution of the moving
APs. This assumption makes it easy to analyze the performance of the
VNETs owing to the tractability of Poisson point process.

Considering uncertain distributions of the vehicles and varied communication
demands in the real scenario, the independent random selection will
not be the better choice than cooperative selection schemes for moving
APs. In cooperative selection schemes, vehicles exchange information
with other vehicles to decide or elect moving APs. According to the
selection manner for the moving APs, the cooperative selection schemes
for moving APs can be divided into two categories that are centralized
selection schemes and decentralized selection schemes. For the centralized
selection schemes, a centered controller of the global or local area
collects all the necessary information from vehicles and then decides
which vehicles are chosen as the moving APs. One of the feasible methods
to realize a centralized selection scheme is to form vehicles as local
moving software-defined cloudlets. If a software-defined cloudlet
controller is deployed on a vehicle, it will be responsible for assigning
local moving APs, and act as a moving AP itself as well. On the contrary,
in decentralized selection schemes, all the candidate vehicles exchange
information and operate equally to select moving APs. The selection
will be organized in a manner of distributed election or broadcasting
competition. Apparently, there are high overheads for exchanging information
while selecting the moving APs, which also bring a great challenge
for the strict latency requirement in VNETs. One of the possible approaches
to reduce the communication overheads is the RSU-assisted selection.
The RSUs, which are connected to each other by high-speed backbone
networks, can exchange information efficiently so that they can possess
the location and velocity information of vehicles.

Whichever cooperative selection scheme is chosen, there are many selection
strategies available for adopting. For simple examples, in this article,
we consider the real locations of vehicles, and then evenly choose
the moving APs over the balanced distance or an equal number of mediate
vehicles between two adjacent moving APs. To be specific, we investigate
the following three kinds of simple selection strategies.
\begin{enumerate}
\item Independent random decision strategy. In this strategy, every vehicle
independently decides if or not it will become an AP according to
a given probability. It is a simple strategy without any extra communication
overhead.
\item Sequence-based selection strategy. Vehicles are assigned sequential
number according to the order of their locations on the road. The
moving APs will be selected from the vehicles according to a given
selecting ratio. For instance, if the selecting ratio is $0.1$, one
moving AP will be sequentially selected from every ten vehicles. This
strategy is simple and requires low communication overhead for counting
the sequential numbers by exchanging information with the adjacent
vehicles. This approach will help precisely balance the load of the
moving APs, and improve the connectivity for dense vehicles in a local
area.
\item Distance-based selection strategy. In this strategy, the moving APs
are selected spatially evenly over distance, so that the coverage
ranges of every moving AP are comparable. The goal of the strategy
is to improve the coverage range on the road. The distance-based selection
strategy requires higher communication overheads than the sequence-based
one, because it needs to exchange location data besides sequential
information.
\end{enumerate}
The above strategies are simple yet workable. They can be performed
by the vehicles in a distributed manner, and also can be carried out
by a local software-defined cloudlet controller. Some complicated
strategies, such as the auction-based selection and distributed competition,
can be further considered as well.

\subsection{Connectivity to Moving APs}

The connectivity of 5G moving-AP-based cell-less communications can
be evaluated from three aspects, including the connectivity probability
that vehicle users can access to moving APs, the probability that
vehicle users can access to the network by D2D links, and the probability
that vehicle users can access to nearby fixed 5G BSs. In this article,
we focus on the connectivity probability that vehicle users can access
to moving APs.

We set up an illustrative simulation scenario for performance analysis
where the vehicles on a $10\,\mathrm{km}$ of road follow a Poisson
point process with the density of $0.02\,\mathrm{m}^{-1}$, and their
velocities follow a uniform distribution within $[50,80]\,\mathrm{km/h}$.
The velocity of every vehicle keeps constant during the simulation.
It is assumed that the communication channels follow independent Rayleigh
fading with a pathloss exponent $4$, the transmission power of the
moving APs is $2\,\mathrm{W}$, the noise power received at vehicle
users is $-100\,\mathrm{dBw}$. In this simulation, a vehicle user
uses received SINR to choose a moving AP to communicate with in the
non-cooperative mode, or two cooperative APs in the cooperative mode.
The downlink connectivity probability of a vehicle user accessing
to cooperative moving APs is compared to the connectivity probability
of the vehicle user accessing to a non-cooperative moving AP. The
impacts of different moving AP selecting probabilities and different
selection strategies are evaluated as well.

\begin{figure}[tbh]
\begin{centering}
\includegraphics{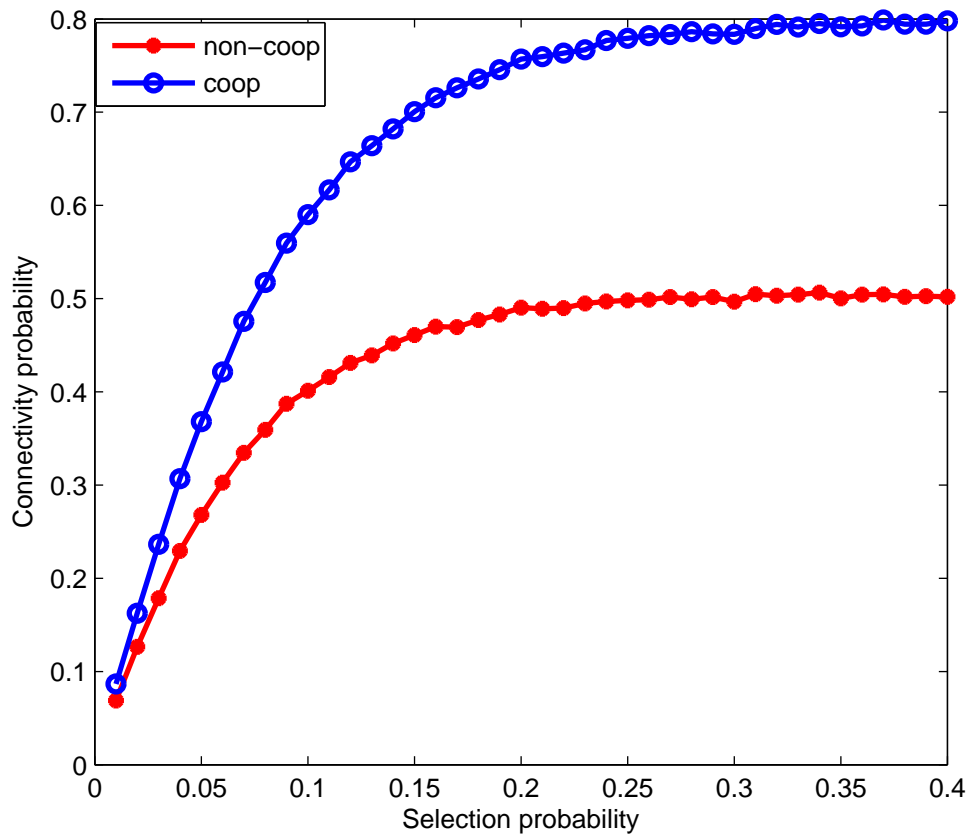}
\par\end{centering}
\caption{Connectivity with respect to the moving AP selection probability\label{fig:Connectivity-over-selec}}
\end{figure}

Figure \ref{fig:Connectivity-over-selec} shows the relationship between
the connectivity probabilities and the moving AP selection probability,
as independent random decision strategy is adopted. It can be seen
that for accessing to both cooperative APs and a non-cooperative AP,
the connectivity probabilities increase as the moving AP selection
probability increases. That is, a higher density of moving APs makes
it easier for the vehicle users to access to the network. Furthermore,
it is observed that the connectivity probabilities approach a floor,
where it is not always beneficial to increase the moving AP selection
probability. It means that the density of moving APs needs to be properly
designed to reach a balance between the connectivity and the corresponding
energy consumption and costs.

\begin{figure}[tbh]
\begin{centering}
\includegraphics[width=0.9\textwidth]{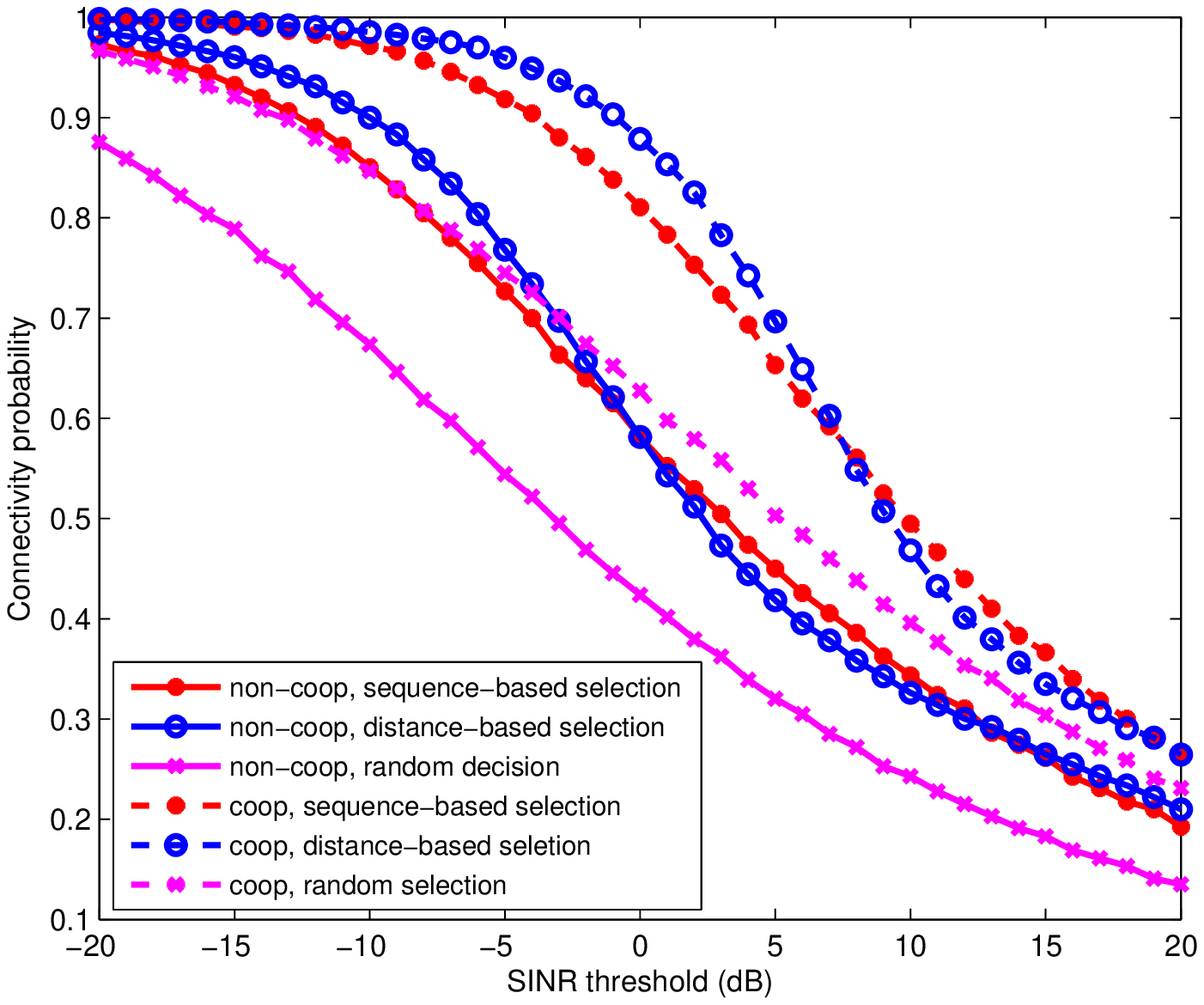}
\par\end{centering}
\caption{Connectivity probabilities of accessing to moving APs under different
selection strategies\label{fig:Connectivity}}
\end{figure}

Figure \ref{fig:Connectivity} shows the relationship between the
downlink connectivity probability that a vehicle user can directly
access to a moving AP, or APs by cooperative communications, and the
SINR threshold of the reception at the vehicle user. In the simulation,
we compare three kinds of the moving AP selection strategies, including
the independent random decision, sequence-based selection, and distance-based
selection. In all the scenarios for these three strategies, we choose
the parameters so that there are the same number of the vehicles,
that is 200 vehicles, in these three scenarios. It can be seen that
the connectivity probabilities of accessing to moving APs in the cooperative
manner are significantly higher than that in non-cooperative ones.
The results also indicate that the distance-based moving AP selection
strategy and the sequence-based selection strategy outperform the
independent random decision strategy. The distance-based strategy
is aimed at the extending the spatial coverage along the road, and
the sequence-based one is designed to balance load and leads to more
moving APs selected among locally dense vehicle users than sparse
users. Selection depending on the real distribution of the vehicles
is the reason that both of them are better than the selection strategy
by totally random decision.

\subsection{Latency in Terms of Relay Hops}

Latency performance is crucial in VNETs, especially for vehicle secure
communications. The latency in 5G moving-AP-based cell-less communications
is affected by many factors including the number of hops from peer
to peer, relay forwarding time, outage probability and time, and other
processing time incurred along the communication path. In this article,
we focus on the number of hops in the communication path between two
end vehicle users, which significantly affects the latency.

\begin{figure}[tbh]
\begin{centering}
\includegraphics[width=0.8\textwidth]{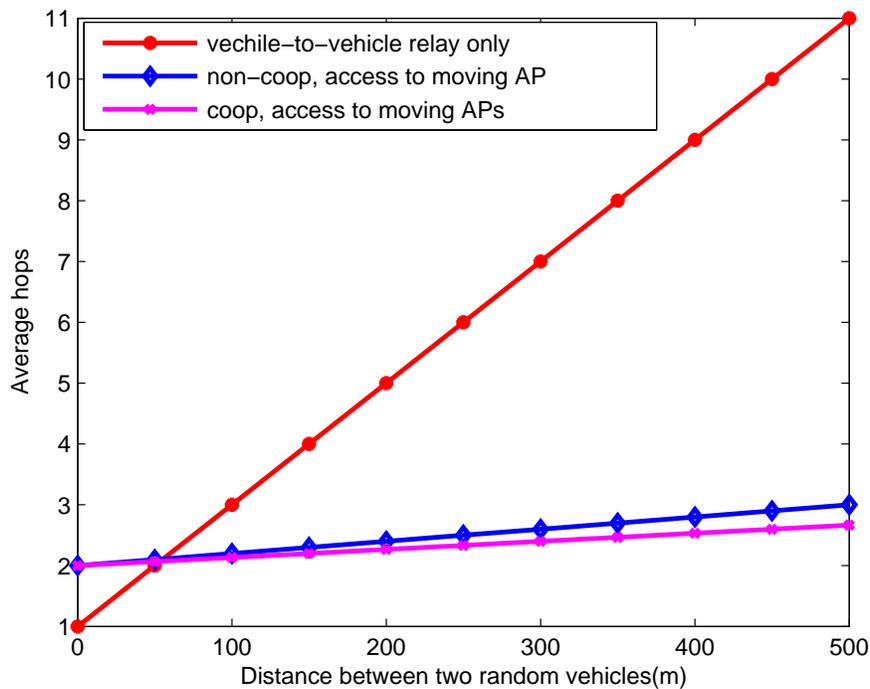}
\par\end{centering}
\caption{The number of hops along the transmission path between two random
vehicle users\label{fig:Hops}}
\end{figure}

Following the same configurations as in Figure \ref{fig:Connectivity-over-selec}
and \ref{fig:Connectivity}, Figure \ref{fig:Hops} shows the relationship
between the average number of hops between two vehicle peers and their
distance under both cases of using moving-AP-based access scheme and
using only multi-hop D2D links. In the moving-AP-based access scheme,
we compare the average hops for the case of accessing to cooperative
moving APs to the case of accessing to non-cooperative moving APs.
It can be seen that with an increase in the distance between peers,
the average number of hops of the moving-AP-based scheme is much smaller
than the D2D-only-based one. The average number of hops between end
users by accessing to cooperative moving AP is always lower than accessing
to non-cooperative moving APs. Furthermore, when the distance between
two communicating vehicle peers is less than $500\,\mathrm{m}$, it
is observed that the average number of hops of the moving-AP-based
scheme between them is between $2$ and $3$. It is owing to the various
transmission models, especially in case (b), as discussed in Figure
\ref{fig:transmission_path}.

\section{Future Challenges}

Although performance gains in terms of latency and connectivity can
be expected, the 5G moving-AP-based cell-less VNETs also face many
challenges. Some future challenges are listed below.
\begin{enumerate}
\item In the future 5G VNETs, the interference and routing issues caused
by the heterogeneity of the 5G networks will be major challenges.
Using 5G technologies in VNETs will lead to interference with the
common 5G networks in the same areas. Considering the heterogeneous
5G networks consist of various radio technologies, when the vehicle
users move on the road and communicate with the moving APs, they will
encounter fast-changing and complicated interference from the fixed
5G networks near the road. Especially in the future 5G ultra-dense
networks, the mobility of the vehicle users become major concerns
for evaluating and eliminating interference. On the other side, the
moving AP will also cause severe interference to the ordinary users
associated to BSs along the road. Besides interference, the mobility
of the vehicle users and the heterogeneity and ultra-density of 5G
networks can also cause serious routing issues. The transmission path
should be determined over rapid-changing topology and across the different
heterogeneous network tiers. Although the cell-less manner access
alleviates the routing problems, how to choose the optimal transmission
path from the source moving AP to the destination moving AP is still
a performance-critical issue.
\item More efficient backhaul links are required for 5G moving-AP-based
cell-less VNETs. To cooperatively communicate with vehicle users,
the moving APs have to establish stable, high-rate wireless backhaul
links among them. However, due to the movements of vehicle users and
the quick change of the surrounding radio environments, it is not
easy to maintain the quality of the backhaul links. To solve this
problem, some new backhaul management schemes should be proposed to
provide ready-to-use communication links for backup backhaul transmissions.
Efficient prediction on the vehicle movement and communication channel
status will help to create and maintain the backhaul links.
\item The application of new technologies, such as VLC and millimeter wave
transmissions, will bring new technical challenges to 5G moving-AP-based
cell-less VNETs. Intuitively, VLC is suitable for vehicle-to-vehicle
(V2V) communications where modulatable LED lights can be used to replace
the traditional vehicle lights easily, thus providing a much higher
data rate than other rivals \cite{pathak_visible_2015_IEEECommun.Surv.Tutor.}.
However, when many vehicles move along the road, and one of them tries
to communicate with one another using the VLC technology, it very
likely receives the light interference from the vehicles after or
in front of it, and the vehicles running on the middle way of the
communications peers probably block the transmitted signal. Then how
to maintain VLC links over the rapidly changed environment is an important
issue that remains to be investigated in the future. Similarly, although
millimeter wave transmissions can provide significantly high-rate
communication links in line-of-sight situations, they suffer from
noise and propagation loss in air. Therefore, the transmission performance
of millimeter wave communications in such environment remains to be
investigated.
\item Application-awareness is also required in future 5G moving-AP-based
cell-less VNETs. Many vehicular applications tend to send broadcasting
information, such as the traffic accident information that should
be received by all the nearby vehicles as soon as possible, while
some other vehicle applications tend to send unicasting or multicasting
information such as the entertainment data. Depending on the specific
application requirements, the application-aware VNETs would better
provide broadcasting and multicasting services besides the unicasting
one, such that huge duplicated transmissions can be avoided. To realize
such application-aware VNET services, some cross-layer protocols need
to be designed for the future 5G moving-AP-based cell-less VNETs.
\end{enumerate}

\section{Conclusions}

The access network technology is very critical in VNETs, which is
a decisive performance factor of VNETs. With the trend that more and
more V2V communications in VNETs are supported by 5G communication
technologies, it is necessary to adapt the fixed cellular communications
to moving-AP-based cell-less communications. In this article, we propose
a moving-AP-based 5G cell-less VNET scheme, in which fixed BSs are
replaced with on-board moving APs for ease of user access. To strengthen
the connectivity and reliability of the VNET communications, joint
transmission and reception are employed by the moving APs to communicate
cooperatively with the vehicle users. The proposed moving AP scheme
utilizes the cooperative communications between APs to improve the
connectivity in VNETs, whereas a traditional simple moving AP or moving
relay scheme does not use cooperative communications. Taking advantage
of the cooperative communications, the proposed cell-less moving AP
scheme is better than the simple moving relay or moving access schemes
regarding connectivity performance. Illustrative results show that
the connectivity is improved and the number of transmission hops is
reduced significantly.

\section*{Acknowledgment}

The authors would like to acknowledge the support from the International
Science and Technology Cooperation Program of China (Grant No. 2015DFG12580),
the National Natural Science Foundation of China (NSFC) (Grant Nos.
61471180, 61461136004, and 61210002), the Hubei Provincial Department
of Education Scientific research projects (Grant No. B2015188), and
a grant from Wenhua College (No. 2013Y08).

\bibliographystyle{IEEEtran}
\bibliography{IEEEabrv,MWC2016}

\begin{thebibliography}{10}
\providecommand{\url}[1]{#1}
\csname url@samestyle\endcsname
\providecommand{\newblock}{\relax}
\providecommand{\bibinfo}[2]{#2}
\providecommand{\BIBentrySTDinterwordspacing}{\spaceskip=0pt\relax}
\providecommand{\BIBentryALTinterwordstretchfactor}{4}
\providecommand{\BIBentryALTinterwordspacing}{\spaceskip=\fontdimen2\font plus
\BIBentryALTinterwordstretchfactor\fontdimen3\font minus
  \fontdimen4\font\relax}
\providecommand{\BIBforeignlanguage}[2]{{%
\expandafter\ifx\csname l@#1\endcsname\relax
\typeout{** WARNING: IEEEtran.bst: No hyphenation pattern has been}%
\typeout{** loaded for the language `#1'. Using the pattern for}%
\typeout{** the default language instead.}%
\else
\language=\csname l@#1\endcsname
\fi
#2}}
\providecommand{\BIBdecl}{\relax}
\BIBdecl

\bibitem{amadeo_information-centric_2016_IEEECommun.Mag.}
M.~Amadeo, C.~Campolo, and A.~Molinaro, ``Information-centric networking for
  connected vehicles: A survey and future perspectives,'' \emph{IEEE Commun.
  Mag.}, vol.~54, no.~2, pp. 98--104, Feb. 2016.

\bibitem{liu_mobile_2017}
H.~Liu, F.~Eldarrat, H.~Alqahtani, A.~Reznik, X.~d. Foy, and Y.~Zhang, ``Mobile
  {Edge} {Cloud} {System}: {Architectures}, {Challenges}, and {Approaches},''
  \emph{IEEE Syst. J.}, {DOI:} 10.1109/JSYST.2017.2654119, 2017.

\bibitem{zheng_heterogeneous_2015_IEEECommun.Surv.Tutor.}
K.~Zheng, Q.~Zheng, P.~Chatzimisios, W.~Xiang, and Y.~Zhou, ``Heterogeneous
  {{Vehicular Networking}}: {{A Survey}} on {{Architecture}}, {{Challenges}},
  and {{Solutions}},'' \emph{IEEE Commun. Surv. Tutor.}, vol.~17, no.~4, pp.
  2377--2396, Fourthquarter 2015.

\bibitem{salahuddin_software-defined_2015}
M.~A. Salahuddin, A.~Al-Fuqaha, and M.~Guizani, ``Software-{Defined}
  {Networking} for {RSU} {Clouds} in {Support} of the {Internet} of
  {Vehicles},'' \emph{IEEE Internet Things J.}, vol.~2, no.~2, pp. 133--144,
  Apr. 2015.

\bibitem{liao_software_2015_MobileNetwAppl}
L.~Liao, M.~Qiu, and V.~C.~M. Leung, ``\BIBforeignlanguage{en}{Software
  {{Defined Mobile Cloudlet}}},'' \emph{\BIBforeignlanguage{en}{Mobile Networks
  and Applications}}, vol.~20, no.~3, pp. 337--347, May 2015.

\bibitem{vinel_3gpp_2012_IEEEWirel.Commun.Lett.}
A.~Vinel, ``{{3GPP LTE Versus IEEE}} 802.11p/{{WAVE}}: {{Which Technology}} is
  {{Able}} to {{Support Cooperative Vehicular Safety Applications}}?''
  \emph{IEEE Wirel. Commun. Lett.}, vol.~1, no.~2, pp. 125--128, Apr. 2012.

\bibitem{yu_optimal_2016_IEEETrans.Veh.Technol.}
R.~Yu, J.~Ding, X.~Huang, M.~T. Zhou, S.~Gjessing, and Y.~Zhang, ``Optimal
  {{Resource Sharing}} in {{5G}}-{{Enabled Vehicular Networks}}: {{A Matrix
  Game Approach}},'' \emph{IEEE Trans. Veh. Technol.}, vol.~65, no.~10, pp.
  7844--7856, Oct. 2016.

\bibitem{laiyemo_transmission_2016}
A.~O. Laiyemo, H.~Pennanen, P.~Pirinen, and M.~Latva-aho, ``Transmission
  {{Strategies}} for {{Throughput Maximization}} in {{High}}-{{Speed}}-{{Train
  Communications}}: {{From Theoretical Study}} to {{Practical Algorithms}},''
  \emph{IEEE Trans. Veh. Technol.}, vol.~66, no.~4, pp. 2997--3011, Apr. 2017.

\bibitem{jangsher_backhaul_2016}
S.~Jangsher and V.~O.~K. Li, ``Backhaul {{Resource Allocation}} for
  {{Existing}} and {{Newly Arrived Moving Small Cells}},'' \emph{IEEE Trans.
  Veh. Technol.}, vol.~66, no.~4, pp. 3211--3219, Apr. 2017.

\bibitem{patra_improving_2016}
M.~Patra, R.~Thakur, and C.~S.~R. Murthy, ``Improving {{Delay}} and {{Energy
  Efficiency}} of {{Vehicular Networks Using Mobile Femto Access Points}},''
  \emph{IEEE Trans. Veh. Technol.}, vol.~66, no.~2, pp. 1496--1505, Feb. 2017.

\bibitem{han_5g_2016_IEEECommun.Mag.}
T.~Han, X.~Ge, L.~Wang, K.~S. Kwak, Y.~Han, and X.~Liu, ``{{5G Converged
  Cell}}-{{Less Communications}} in {{Smart Cities}},'' \emph{IEEE Commun.
  Mag.}, vol.~55, no.~3, pp. 44--50, Mar. 2017.

\bibitem{feteiha_enabling_2015}
M.~F. Feteiha and H.~S. Hassanein, ``Enabling {Cooperative} {Relaying} {VANET}
  {Clouds} {Over} {LTE}-{A} {Networks},'' \emph{IEEE Trans. Veh. Technol.},
  vol.~64, no.~4, pp. 1468--1479, Apr. 2015.

\bibitem{fodor_overview_2016_IEEEAccess}
G.~Fodor, S.~Roger, N.~Rajatheva, S.~B. Slimane, T.~Svensson, P.~Popovski,
  J.~M. B.~D. Silva, and S.~Ali, ``An {{Overview}} of {{Device}}-to-{{Device
  Communications Technology Components}} in {{METIS}},'' \emph{IEEE Access},
  vol.~4, pp. 3288--3299, 2016.

\bibitem{ge_5g_2014_IEEENetwork}
X.~Ge, H.~Cheng, M.~Guizani, and T.~Han, ``{{5G}} wireless backhaul networks:
  {{Challenges}} and research advances,'' \emph{IEEE Netw.}, vol.~28, no.~6,
  pp. 6--11, 2014.

\bibitem{pathak_visible_2015_IEEECommun.Surv.Tutor.}
P.~H. Pathak, X.~Feng, P.~Hu, and P.~Mohapatra, ``Visible {{Light
  Communication}}, {{Networking}}, and {{Sensing}}: {{A Survey}}, {{Potential}}
  and {{Challenges}},'' \emph{IEEE Commun. Surv. Tutor.}, vol.~17, no.~4, pp.
  2047--2077, Fourthquarter 2015.

\end{thebibliography}

\vspace*{2\baselineskip}
\begin{IEEEbiographynophoto}{Lijun Wang}
 {[}M'16{]} (wanglijun@whu.edu.cn) is pursuing her Ph.D. degree with
the School of Electronic Information, Wuhan University, Wuhan, China.
She is currently an associate professor with the Faculty of Information
Science and Technology, Wenhua College, Wuhan, China. Her research
interests include wireless communications, and multimedia communications.
\end{IEEEbiographynophoto}

\vspace*{-2\baselineskip}
\begin{IEEEbiographynophoto}{Tao Han}
 {[}M'13{]} (hantao@hust.edu.cn) received his Ph.D. degree in information
and communication engineering from Huazhong University of Science
and Technology (HUST), Wuhan, China in December, 2001. He is currently
an associate professor with the School of Electronic Information and
Communications, HUST. His research interests include wireless communications,
multimedia communications, and computer networks. He is currently
serving as an Area Editor for the \emph{EAI Endorsed Transactions
on Cognitive Communications}.
\end{IEEEbiographynophoto}

\vspace*{-2\baselineskip}
\begin{IEEEbiographynophoto}{Qiang Li}
 {[}M'16{]} (qli\_patrick@hust.edu.cn) received his B.Eng. degree
from the University of Electronic Science and Technology of China
(UESTC), China, in 2007, and his Ph.D. degree from Nanyang Technological
University (NTU), Singapore, in 2011. From 2011 to 2013 he was a research
fellow with Nanyang Technological University. Since 2013 he has been
an associate professor with Huazhong University of Science and Technology,
China. His research interests include next generation mobile communications,
software-defined networking, full-duplex techniques, and wireless
cooperative communications.
\end{IEEEbiographynophoto}

\vspace*{-2\baselineskip}
\begin{IEEEbiographynophoto}{Jia Yan}
 (yanjia@whu.edu.cn) received his B.S. and Ph.D. degrees in the School
of Electronic and Information from Wuhan University, Wuhan, China,
in 2005 and 2010. Then he was a postdoctoral research fellow till
2014, and now he is a lecturer with the Department of Electrical Engineering,
Wuhan University. His research interests include visual tracking by
detection and object recognition.
\end{IEEEbiographynophoto}

\vspace*{-2\baselineskip}
\begin{IEEEbiographynophoto}{Xiong Liu}
 (m201571774@hust.edu.cn) received his Bachelor\textquoteright s
degree in electronic information and communication from Huazhong University
of Science and Technology, Wuhan, China, in 2015, where he is currently
pursuing his Master\textquoteright s degree. His research interests
include vehicular networks, non-orthogonal multiple access, and cognitive
radio. 
\end{IEEEbiographynophoto}

\vspace*{-2\baselineskip}
\begin{IEEEbiographynophoto}{Dexiang Deng}
 (ddx@whu.edu.cn) received his B.S. and M.S. degrees from Wuhan Institute
of Surveying and Mapping, Wuhan, China, in 1982 and 1988. From 1999
to 2000, he was a research fellow with the Zurich Polytechnic University,
Switzerland. Currently, he is a professor with the School of Electronic
and Information, Wuhan University. His research interests include
pattern recognition, multimedia technology, and spatial image processing.
\end{IEEEbiographynophoto}

\end{document}